\documentclass[12pt,english]{iopart}
\usepackage[T1]{fontenc}
\usepackage[latin9]{inputenc}
\usepackage{babel}
\usepackage{graphicx}
\usepackage{cite}
\usepackage{float}
\usepackage{subfig}
\usepackage{caption}

\begin{document}

\title[Superfluid phases of spin-1 bosons in cubic optical lattice]{Superfluid phases of spin-1 bosons in cubic optical lattice}

\author{Mohamed Mobarak}
\address{Insitute f\"ur Theoretische Physik, Freie Universit\"at Berlin, Arnimallee 14, 14195 Berlin, Germany}
\author{Axel Pelster}

\address{Fachbereich Physik und Forschungszentrum OPTIMAS, Technische Universit\"at Kaiserslautern, 67663 Kaiserslautern,Germany}
\address{Hanse-Wissenschaftskolleg, Lehmkuhlenbusch 4, D-27733 Delmenhorst, Germany}
\ead{axel.pelster@physik.uni-kl.de}

\begin{abstract}
We analyze theoretically the emergence of different superfluid phases of spin-1 bosons in a three-dimensional
cubic optical lattice by generalizing the recently developed Ginzburg-Landau theory for the Bose-Hubbard
model to a spinor Bose gas. In particular at zero temperature, our theory distinguishes within its validity
range between various superfluid phases for an anti-ferromagnetic interaction with an external magnetic
field.  In addition, we determine that the superfluid-Mott insulator phase transition is of second order
and that the transitions between the respective superfluid phases with anti-ferromagnetic interaction
can be both of first and second order.
\end{abstract}

\section{Introduction}
In recent years optical lattices have become a major research topic within the realm of ultracold quantum
gases as they offer the perspective to simulate condensed matter physics under well-controlled conditions \cite{key-90,key-37}.
Most prominently, the quantum phase transition between a superfl{}uid (SF) and a Mott-insulating (MI)
phase of a spinless Bose gas loaded in a periodic optical potential has been experimentally observed
by increasing the lattice depth \cite{key-34,key-35}. All properties of this quantum phase transition
are captured by the underlying Bose-Hubbard Hamiltonian \cite{key-33,key-50,key-36} for which different
analytical solution methods have been worked out \cite{key-22,key-18,key-23,key-3,key-24,key-25,key-30}
and high-precision Monte Carlo studies have been performed \cite{key-27,key-28}. Furthermore, extensions
of the Bose-Hubbard model have been investigated, which cover for instance, superlattices \cite{key-31},
Bose-Fermi mixtures \cite{key-58,key-59,key-60,key-61}, quantum simulations like entanglement of atoms
or quantum teleportation \cite{key-57} and disorder \cite{key-51,key-53-1,key-52,key-37}.

Preparing experimentally a spin-1 Bose-Einstein condensation (BEC)  of $^{23}$Na or $^{87}$Rb atoms in an optical trap the atomic
spin degrees of freedom are not frozen due to the electric dipole force between atoms and the electric
field of a laser beam \cite{key-95,key-86-1}. This experimental realization of an optically trapped
BEC opened a new window to study also various phenomena of spinor Bose gases loaded in an optical lattice.
For instance, they offer the possibility of studying strongly correlated states, for example the coherent
collisional spin dynamics in an optical lattice was measured in Ref.~\cite{key-42-2} and the $^{87}$Rb
scattering lengths for $F=1$ and $F=2$ were determined in Ref.~\cite{key-42-3}. In particular, combining
the spin degree of freedom with various types of interactions and with different lattice geometries offers
the prospect to realize a plethora of superfluid phases with magnetic properties. A first tentative step
in this direction was the loading of $^{87}$Rb in a frustrated triangular lattice \cite{key-53}. Despite
these initial promising investigations, spinor Bose gases in optical lattice seem experimentally to be
so challenging that no further detailed experiments have been performed. 

Theoretical progress in the study of spinor Bose gases in an optical lattice was made by Refs.~\cite{key-29,key-6}.
In the case of the anti-ferromagnetic interaction of $^{23}$Na the location of the SF-MI transition
and several properties for SF and MI phases for spin-1 bosons were determined without external magnetic
field at zero temperature. In particular, they found that the superfluid transition occurs into a polar
spin-0 state \cite{key-6} and the SF phase represents a polar state with zero spin expectation value.
On the other hand, the effect of a non-vanishing external magnetic field upon the SF-MI transition was
determined within a mean-field approximation in Refs.~\cite{key-60-1,key-15}. In addition,
it was also shown in Refs.~\cite{key-60-1,key-15} that the superfluid transition occurs from Mott insulating phase into either  a polar
spin-1 or spin-(-1) state but it was not investigated which other phases  might emerge deep in the superfluid phase.

In this Letter we study the effect of the magnetic field on the emergence of superfluid phases of spin-1 bosons. We show that new superfluid phases can emerge due to the interplay of the
anti-ferromagnetic interaction of spin-1 bosons and an external magnetic field in a three-dimensional cubic optical lattice at zero temperature. To this end, we extend the Ginzburg-Landau theory developed
in Refs.~\cite{key-3,key-23} from the spin-0 to the spin-1 Bose-Hubbard model. In the grand-canonical
ensemble the underlying Hamiltonian can be decomposed according to $\hat{H}_{\rm{BH}}=\hat{H}^{\left(0\right)}+\hat{H}^{\left(1\right)}$,
\cite{key-6,key-29} where the local part $\hat{H}^{\left(0\right)}=\sum_{i}\hat{H}_{i}^{\left(0\right)}$
reads
\begin{eqnarray} 
\hat{H}_{i}^{\left(0\right)} & =\frac{U_{0}}{2}\hat{n}_{i}(\hat{n}_{i}-1)+\frac{U_{2}}{2}(\hat{\mathbf{S}}_{i}^{2}-2\hat{n}_{i})-\mu\hat{n}_{i}-
\eta\hat{S}_{iz},\label{eq:1}
\end{eqnarray}
whereas the bilocal part is given by 
\begin{eqnarray}
\hat{H}^{\left(1\right)} & =-J\sum_{\left\langle i,j\right\rangle }\sum_{\alpha}\hat{a}_{i\alpha}^{\dagger}\hat{a}_{j\alpha}.
\end{eqnarray}
Here $\mu$ and $\eta$ denote the chemical potential and the external magnetic field, respectively.
Furthermore,\emph{ $J$} represents the hopping matrix element between adjacent sites \emph{$i$ }and
\emph{$j$} with $\left\langle i,j\right\rangle $ indicating the summation over all nearest neighbor
sites and \emph{$U{}_{0}$ $\left(U_{2}\right)$} stands for the on-site spin-independent (dependent)
interaction strength between bosons. Additionally, $\hat{a}_{i\alpha}$ $\left(\hat{a}_{i\alpha}^{\dagger}\right)$
is the annihilation (creation) operator at site $i$ with hyperfine spin $\alpha\in\left\{ -1,0,1\right\} $
which determines the total atom number operator at site $i$ via $\hat{n}_{i}=\sum_{\alpha}\hat{a}_{i\alpha}^{\dagger}\hat{a}_{i\alpha}$
and the spin operator at site $i$ according to $\hat{\mathbf{\mathbf{S}}}_{i}=\sum_{\alpha,\beta}\hat{a}_{i\alpha}^{\dagger}\mathbf{\,\mathbf{F}}_{\alpha\beta}\,\hat{a}_{i\beta}$
with the spin-1 matrices $\mathbf{F_{\alpha\beta}}$. Since the operators $\hat{\mathbf{S}}_{i}^{2}$,
$\hat{S}_{iz}$ and $\hat{n}_{i}$ commute with each other, their eigenvalue problems are solved by the
same eigenvectors: $\hat{\mathbf{S}}_{i}^{2}\left|\mathit{S}_{i},m_{i},n_{i}\right\rangle =\mathit{S}_{i}(\mathit{S}_{i}+1)\left|S_{i},m_{i},n_{i}\right\rangle ,$
$\hat{S}_{iz}\left|S_{i},m_{i},n_{i}\right\rangle =m_{i}\left|S_{i},m_{i},n_{i}\right\rangle $ and $\hat{n}_{i}\left|S_{i},m_{i},n_{i}\right\rangle =n_{i}\left|S_{i},m_{i},n_{i}\right\rangle $
where $S_{i}+n_{i}$ must be an even number \cite{key-42,key-29,key-6,key-60-3}. Thus, the eigenvalue problem
of the local Hamiltonian (\ref{eq:1}) is given by 
\begin{eqnarray}
\hat{H}_{i}^{\left(0\right)}\left|S_{i},m_{i},n_{i}\right\rangle  & =E_{S_{i},m_{i},n_{i}}^{\left(0\right)}\left|S_{i},m_{i},n_{i}\right\rangle ,
\end{eqnarray}
where the energy eigenvalues are defined as 
\begin{eqnarray}
E_{S_{i},m_{i},n_{i}}^{\left(0\right)}= & \frac{U_{0}}{2}n_{i}(n_{i}-1)+\frac{U_{2}}{2}\bigl[S_{i}(\mathit{S}_{i}+1)-2n_{i}\bigr]
 & -\mu n_{i}-\eta m_{i}.\label{eq:3}
\end{eqnarray}

In order to artificially break the underlying $U(1)$ symmetry of the Hamiltonian $\hat{H}_{\rm{BH}}$,
we follow Refs.~\cite{key-3,key-23} and generalize the usual field-theoretic approach for describing
classical phase transitions \cite{key-8,key-9} to the realm of quantum phase transitions. Thus, we couple
artificial source currents $j_{i\alpha}(\tau),j_{i\alpha}^{*}(\tau)$ to the operators $\hat{a}_{i\alpha}^{\dagger}$
and $\hat{a}_{i\alpha}$

\begin{eqnarray}
\hat{H}_{\rm{BH}}(\tau)= & \hat{H}_{\rm{BH}}+\sum_{i}\sum_{\alpha}\left[j_{i\alpha}^{*}(\tau)\hat{a}_{i\alpha}(\tau)+c.c.\right],\label{eq:3-9-1-1-1}
\end{eqnarray}
yielding a Ginzburg-Landau theory with the spatio-temporal order parameters being defined according to

\begin{equation}
\Psi_{i\alpha}(\tau)=\beta\frac{\delta\mathcal{F}}{\delta j_{i\alpha}^{*}(\tau)}.\label{eq:5-218-1}
\end{equation}
Here the free energy $\mathcal{F}\left[j,j^{*}\right]=-\frac{1}{\beta}\ln\mathcal{Z}\left[j,j^{*}\right]$
with $\beta=1/k_{B}T$ follows from the partition function $\mathcal{Z}\left[j,j^{*}\right]=\rm{Tr }\hat{T}e^{-\int_{0}^{\beta}d\tau\hat{H}_{\mathrm{BH}}(\tau)}$
with the time-ordering operator $\hat{T}$ and the convention $\hbar=1$. We consider Eq.~(\ref{eq:5-218-1})
as a motivation to perform a functional Legendre transformation and define the effective action according
to 
\begin{eqnarray}
\Gamma\left[\Psi,\Psi^{*}\right]=\mathcal{F}\left[j,j^{*}\right]
-\frac{1}{\beta}\sum_{i}\sum_{\alpha}\left[\Psi_{i\alpha}(\tau)j_{i\alpha}^{*}(\tau)+\Psi_{i\alpha}^{*}(\tau)j_{i\alpha}(\tau)\right],\label{eq:4-20}
\end{eqnarray}
where $\Psi_{i\alpha}$ $\left(\Psi_{i\alpha}^{*}\right)$ and $j_{i\alpha}^{*}$ $\left(j_{i\alpha}\right)$
are conjugate variables satisfying the Legendre relations
\begin{eqnarray}
j_{i\alpha}(\tau)=-\beta\frac{\delta\Gamma}{\delta\Psi_{i\alpha}^{*}(\tau)},\quad j_{i\alpha}^{*}(\tau)=-\beta\frac{\delta\Gamma}{\delta\Psi_{i\alpha}(\tau)}.\label{eq:8}
\end{eqnarray}
In order to recover the relevant physical situation the artificial currents $j^{*}$, $j$ should vanish.
Therefore, we obtain from Eq.~(\ref{eq:8}) equations  of motion for determining the equilibrium value
of the order parameter: 
\begin{eqnarray}
\left.\frac{\delta\Gamma}{\delta\Psi_{i\alpha}^{*}(\tau)}\right|_{\Psi=\Psi_{\rm{eq}}}=0,\left.\frac{\delta\Gamma}{\delta\Psi_{i\alpha}(\tau)}\right|_{\Psi=\Psi_{\rm{eq}}}=0.\label{eq:4-22}
\end{eqnarray}
Furthermore, we read off from Eq.~(\ref{eq:4-20}) that evaluating the effective action at the equilibrium
field $\Psi_{\mathrm{eq}}$ recovers the physical grand-canonical free energy: 
\begin{eqnarray}
\Gamma\left[\Psi_{\rm{eq}},\Psi_{\rm{eq}}^{*}\right]=\mathcal{F}\left[0,0\right].\label{eq:4-22-1-1}
\end{eqnarray}
In order to calculate both the free energy $\mathcal{F}$ and the effective action $\Gamma$, we proceed
perturbatively as follows. We decompose the generalized Bose-Hubbard Hamiltonian according to $\hat{H}_{\mathrm{BH}}(\tau)=\hat{H}^{\left(0\right)}+\hat{H}^{\left(1\right)}(\tau)\left[j,j^{*}\right]$,
where the perturbative Hamiltonian in the imaginary-time Dirac interaction picture reads 
\begin{eqnarray}
\hat{H}_{I}^{\left(1\right)}(\tau)\left[j,j^{*}\right]=-J\sum_{\left\langle i,j\right\rangle }\sum_{\alpha}\hat{a}_{i\alpha}^{\dagger}(\tau)\hat{a}_{i\alpha}
(\tau)\nonumber\\
\quad\quad\quad\quad\quad\quad\quad+\sum_{i}\sum_{\alpha}\left[j_{i\alpha}^{*}(\tau)\hat{a}_{i\alpha}(\tau)+j_{i\alpha}(\tau)\hat{a}_{i\alpha}^{\dagger}(\tau)\right].\label{eq:3-9-1}
\end{eqnarray}
With this, we determine the partition function via the Dyson series 
\begin{eqnarray}
\mathcal{Z}= & \mathcal{Z}^{(0)}\Biggl[1+\sum_{n=1}^{\infty}(-1)^{n}\frac{1}{n!}\int_{0}^{\beta}d\tau_{1}\int_{0}^{\beta}d\tau_{2}\cdots\int_{0}^{\beta}d\tau_{n}\:\nonumber \\
 & \times\left\langle \hat{T}\left[\hat{H}_{I}^{(1)}(\tau_{1})\hat{H}_{I}^{(1)}(\tau_{2})\cdots\hat{H}_{I}^{(1)}(\tau_{n})\right]\right\rangle ^{(0)}\Biggr]\label{eq:3-27}
\end{eqnarray}
with $\mathcal{Z}^{(0)}=\mathrm{Tr}e^{-\beta\hat{H}^{(0)}}\,$ and the thermal average defined with respect
to the unperturbed system $\left\langle \bullet\right\rangle ^{(0)}=\mathrm{Tr}\biggl[\bullet\: e^{-\beta\hat{H}^{(0)}}\biggr]/\mathcal{Z}^{(0)}$.
The respective perturbative contributions for $\mathcal{F}$ contain different orders of the hopping
matrix element $J$ and the currents $j$ and $j^{*}$. As we work out a Ginzburg-Landau theory, we restrict
ourselves to the fourth order in the currents. Furthermore, we focus on the leading non-trivial order
in the hopping $J$ which is of first order. Therefore, the free energy functional can be expressed in
terms of imaginary time integrals over sums of products of thermal Green functions. The thermal averages
in Eq.~(\ref{eq:3-27}) can be expressed in terms of $n$-particle Green functions of the unperturbed
system 
\begin{eqnarray}
G_{n}^{(0)}(i_{1}^{\prime}\alpha_{1}^{\prime},\tau_{1}^{\prime};\ldots;i_{n}^{\prime}\alpha_{n}^{\prime},\tau_{n}^{\prime}|i_{1}\alpha_{1},\tau_{1};\ldots;i_{n}\alpha_{n},\tau_{n})\label{eq:3-36-1}\nonumber\\
=\left\langle \hat{T}\left[\hat{a}_{i_{1}^{\prime}\alpha_{1}^{\prime}}^{\dagger}(\tau_{1}^{\prime})\hat{a}_{i_{1}\alpha_{1}}(\tau_{1})\ldots\hat{a}_{i_{n}^{\prime}\alpha_{n}^{\prime}}^{\dagger}(\tau_{n}^{\prime})\hat{a}_{i_{n}\alpha_{n}}(\tau_{n})\right]\right\rangle ^{(0)}.
\end{eqnarray}
In order to calculate the correlation functions in many-body theory, we usually use the Wick theorem
which allows to decompose the $n$-point correlation function (\ref{eq:3-36-1}) into sums of products
of one-point correlation functions \cite{key-11,key-12,key-8,key-9}. However, this theorem is not valid
for the considered system here because the unperturbed Bose-Hubbard Hamiltonian (\ref{eq:1}) contains
terms which are of fourth order in the creation and annihilation operators. Therefore,   we use
the cumulant decomposition for Green function which is based on the locality of $\hat{H}^{\left(0\right)}$
\cite{key-13,key-43}. With this, the unperturbed one- and two-point Green functions are given by
\begin{eqnarray}
G_{1}^{(0)}(i_{1}\alpha_{1},\tau_{1}|i_{2}\alpha_{2},\tau_{2}) & =\delta_{i_{1},i_{2}}\:_{i_{1}}C_{1}^{(0)}(\tau_{1},\alpha_{1}|\tau_{2},\alpha_{2}),\label{eq:3-60}
\end{eqnarray}
and
\begin{eqnarray}
G_{2}^{(0)}(i_{1}\alpha_{1},\tau_{1};i_{2}\alpha_{2},\tau_{2}|i_{3}\alpha_{3},\tau_{3};i_{4}\alpha_{4},\tau_{4})=\nonumber\\                                                                              
\delta_{i_{1},i_{3}}\delta_{i_{2},i_{4}}\delta_{i_{3},i_{4}}\:{}_{i_{1}}C_{1}^{(0)}(\tau_{1},\alpha_{1};\tau_{2},\alpha_{2}|\tau_{3},\alpha_{3};\tau_{4},\alpha_{4})\nonumber \\
+\delta_{i_{1},i_{3}}\delta_{i_{2},i_{4}}\:{}_{i_{1}}C_{1}^{(0)}(\tau_{1},\alpha_{1}|\tau_{3},\alpha_{3})\:{}_{i_{1}}C_{1}^{(0)}(\tau_{2},\alpha_{2}|\tau_{4},\alpha_{4})\nonumber \\
+\delta_{i_{1},i_{4}}\delta_{i_{2},i_{3}}\,{}_{i_{1}}C_{1}^{(0)}(\tau_{1},\alpha_{1}|\tau_{4},\alpha_{4})\:{}_{i_{1}}C_{1}^{(0)}(\tau_{2},\alpha_{2}|\tau_{3},\alpha_{3}).\label{eq:3-61}
\end{eqnarray}
In order to calculate the respective cumulants from combining Eqs.~(\ref{eq:3-36-1})--(\ref{eq:3-61}),
we use for each lattice site the property \cite{key-6,key-10,key-16} 
\begin{eqnarray}
\hat{a}_{\alpha}^{\dagger}\left|S,m,n\right\rangle =M_{\alpha,S,m,n}\left|\mathit{S}+1,m+\alpha,n+1\right\rangle \nonumber \\
\quad\quad\quad\quad\quad\quad +N_{\alpha,S,m,n}\left|S-1,m+\alpha,n+1\right\rangle ,\label{eq:3-31-1}
\end{eqnarray}
\begin{eqnarray}
\hat{a}_{\alpha}\left|S,m,n\right\rangle =O_{\alpha,S,m,n}\left|S+1,m-\alpha,n-1\right\rangle \nonumber \\
\quad\quad\quad\quad\quad\quad+P_{\alpha,S,m,n}\left|S-1,m-\alpha,n-1\right\rangle ,\label{eq:3-32-1}
\end{eqnarray}
where $M_{\alpha,S,m,n}$, $N_{\alpha,S,m,n}$, $O_{\alpha,S,m,n}$ and $P_{\alpha,S,m,n}$ represent
recursively defined matrix elements of creation and annihilation operators. Having calculated the free
energy $\mathcal{F}$ in this way, we perform then the Legendre transformation (\ref{eq:4-20}) to determine
the effective action. In the special case of a stationary equilibrium, which is site-independent due
to homogeneity, the order parameter is given in terms of Matsubara frequencies $\omega_{m}=2\pi m/\beta$:
$\Psi_{i\alpha}^{\rm{eq}}(\omega_{m})=\Psi_{\alpha}\sqrt{\beta}\,\delta_{m,0}\:\:,\:\:\Psi_{i\alpha}^{*\rm{eq}}(\omega_{m})=\Psi_{\alpha}^{*}\sqrt{\beta}\,\delta_{m,0}.$
Thus, the on-site effective potential becomes 
\begin{eqnarray}
\Gamma\left(\Psi_{\alpha},\Psi_{\alpha}^{*}\right)=\mathcal{F}_{0}+\sum_{\alpha}B_{\alpha}\left|\Psi_{\alpha}\right|^{2}
+\sum_{\alpha_{1},\alpha_{2},\alpha_{3},\alpha_{4}}A_{\alpha_{1}\alpha_{2}\alpha_{3}\alpha_{4}}\Psi_{\alpha_{1}}^{*}\Psi_{\alpha_{2}}^{*}\Psi_{\alpha_{3}}\Psi_{\alpha_{4}},\label{eq:18}
\end{eqnarray}
with the Landau coefficients

\begin{eqnarray}
B_{\alpha} & =\frac{1}{a_{2}^{(0)}(\alpha,0)}-zJ,\label{eq:5-284}
\end{eqnarray}
\begin{eqnarray}
A_{\alpha_{1}\alpha_{2}\alpha_{3}\alpha_{4}} & =-\;\frac{\beta a_{4}^{(0)}(\alpha_{1},0;\alpha_{2},0|\alpha_{3},0;\alpha_{4},0)}{4a_{2}^{(0)}(\alpha_{1},0)a_{2}^{(0)}(\alpha_{2},0)a_{2}^{(0)}(\alpha_{3},0)a_{2}^{(0)}(\alpha_{4},0)},\label{eq:5-285}
\end{eqnarray}
where $z=2D$ denotes the coordination number in a $D$-dimensional cubic lattice \cite{key-50-1}. Furthermore,
$a_{2}^{(0)}$ and $a_{4}^{(0)}$ follow from the cumulants but they are not displayed here due to their
complicated and lengthy expressions.

\begin{figure}[t!]
\subfloat[\label{fig:51a}$U_{2}/U_{0}=0.02$.]{\raggedright{}\includegraphics[width=7.5cm]{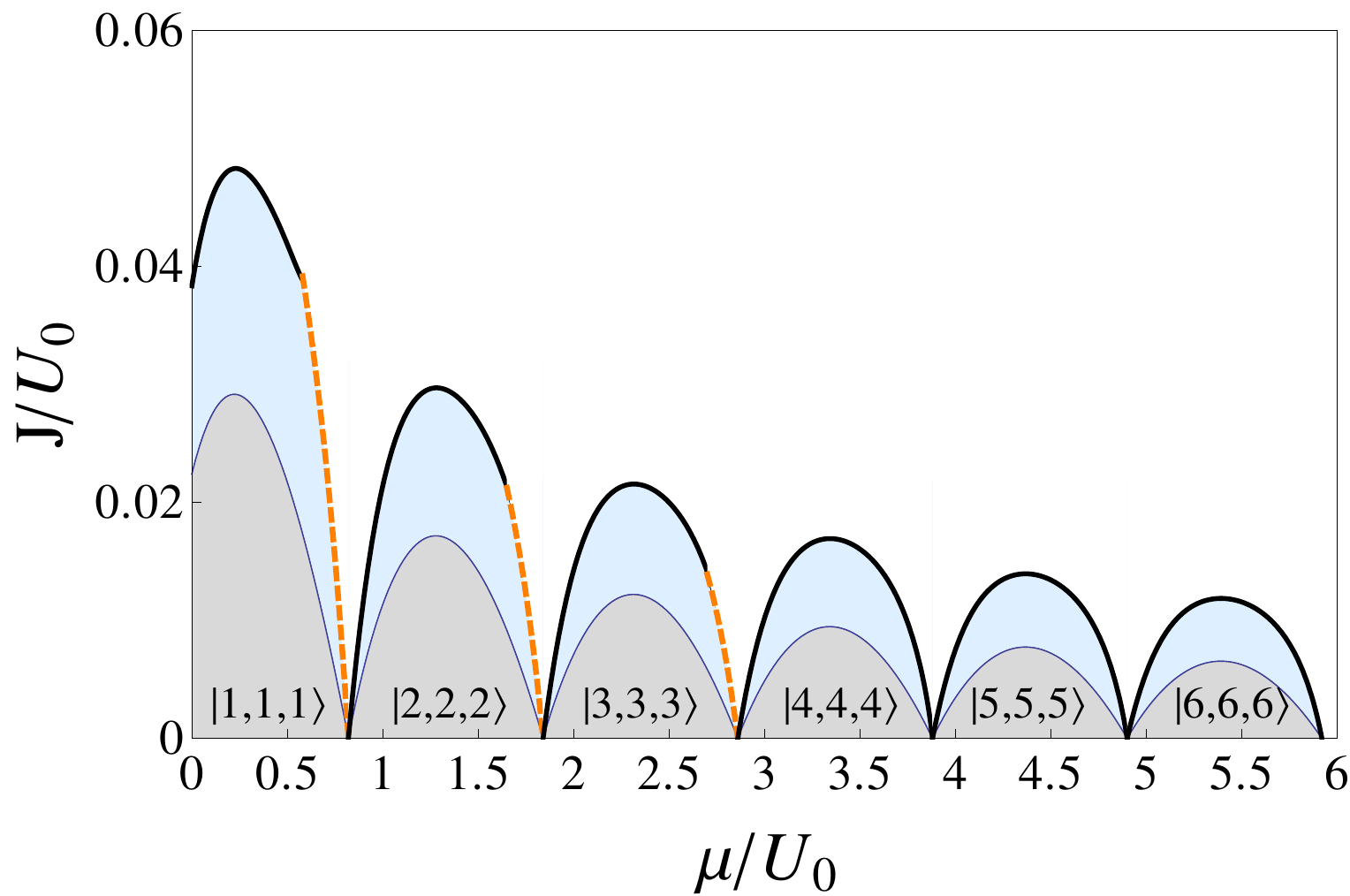}}\hfill{}
\subfloat[\label{fig:51b}$U_{2}/U_{0}=0.04$.]{\raggedright{}\includegraphics[width=7.5cm]{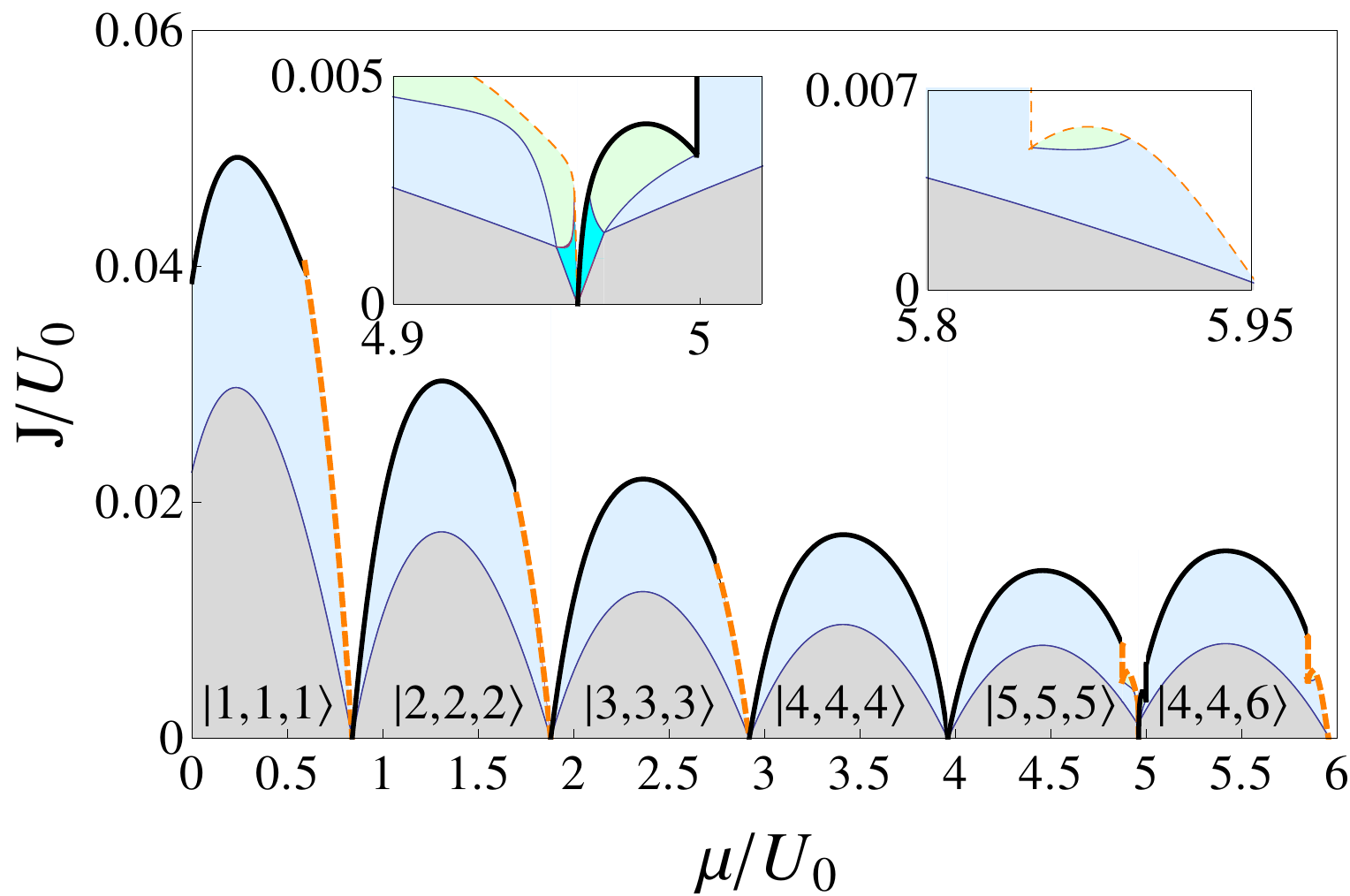}}\hfill{}
\subfloat[\label{fig:51c}$U_{2}/U_{0}=0.05$.]{\raggedright{}\includegraphics[width=7.5cm]{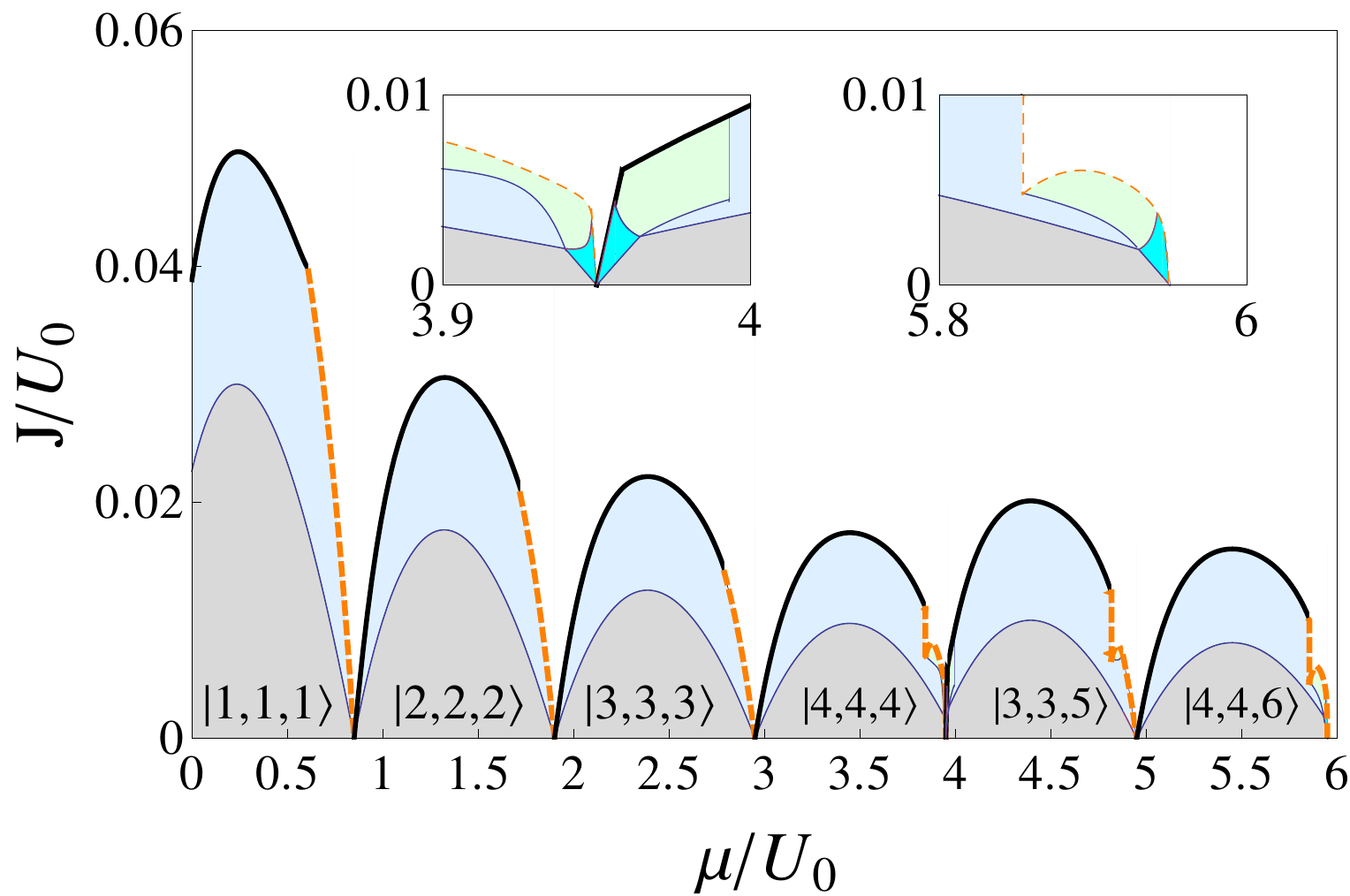}}\hfill{}
\subfloat[\label{fig:51d}$U_{2}/U_{0}=0.07$.]{\raggedright{}\ \ \includegraphics[width=7.5cm]{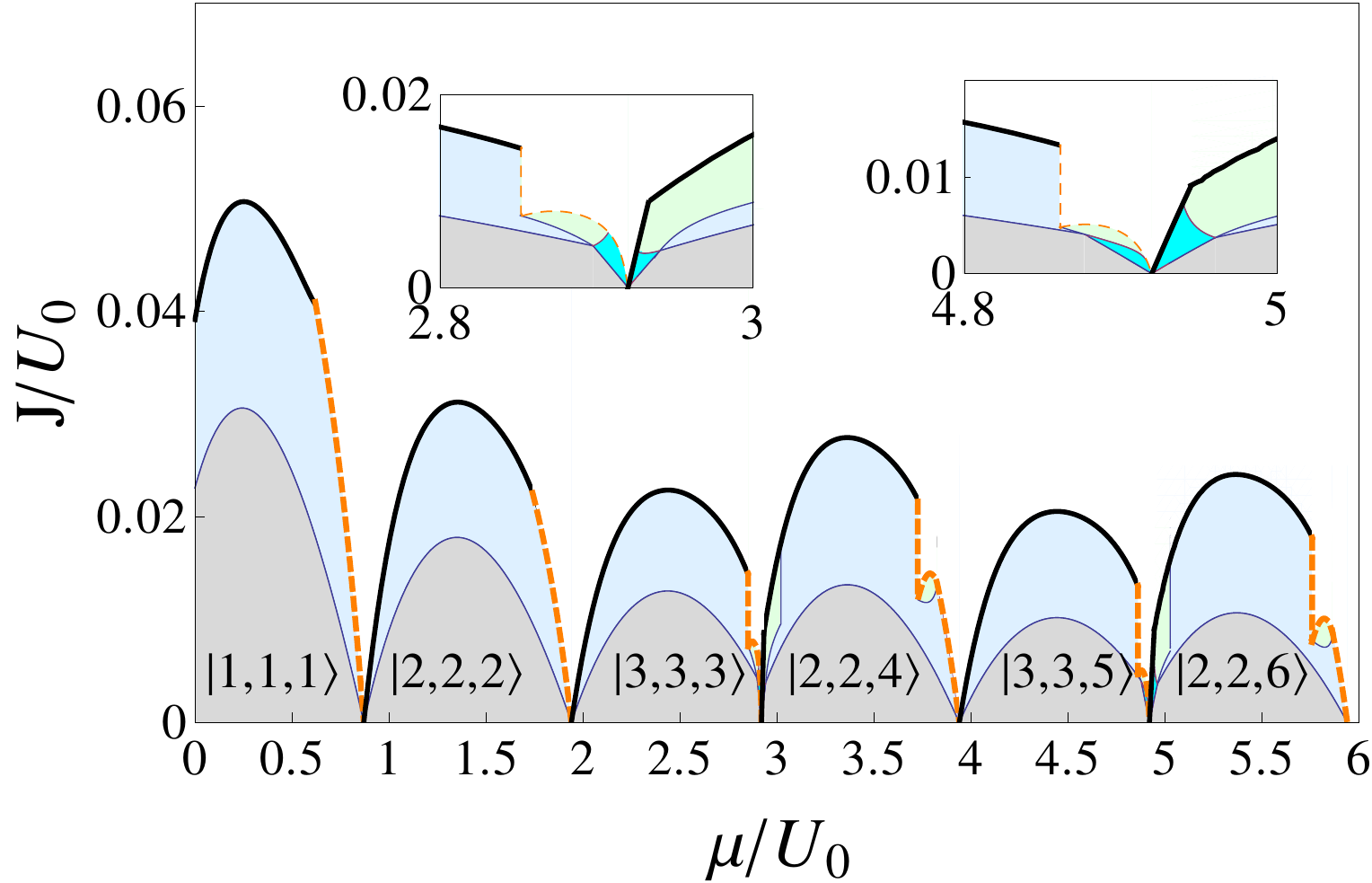}}\hfill{}
\subfloat[\label{fig:51e}$U_{2}/U_{0}=0.1$.]{\raggedright{}\ \ \includegraphics[width=7.5cm]{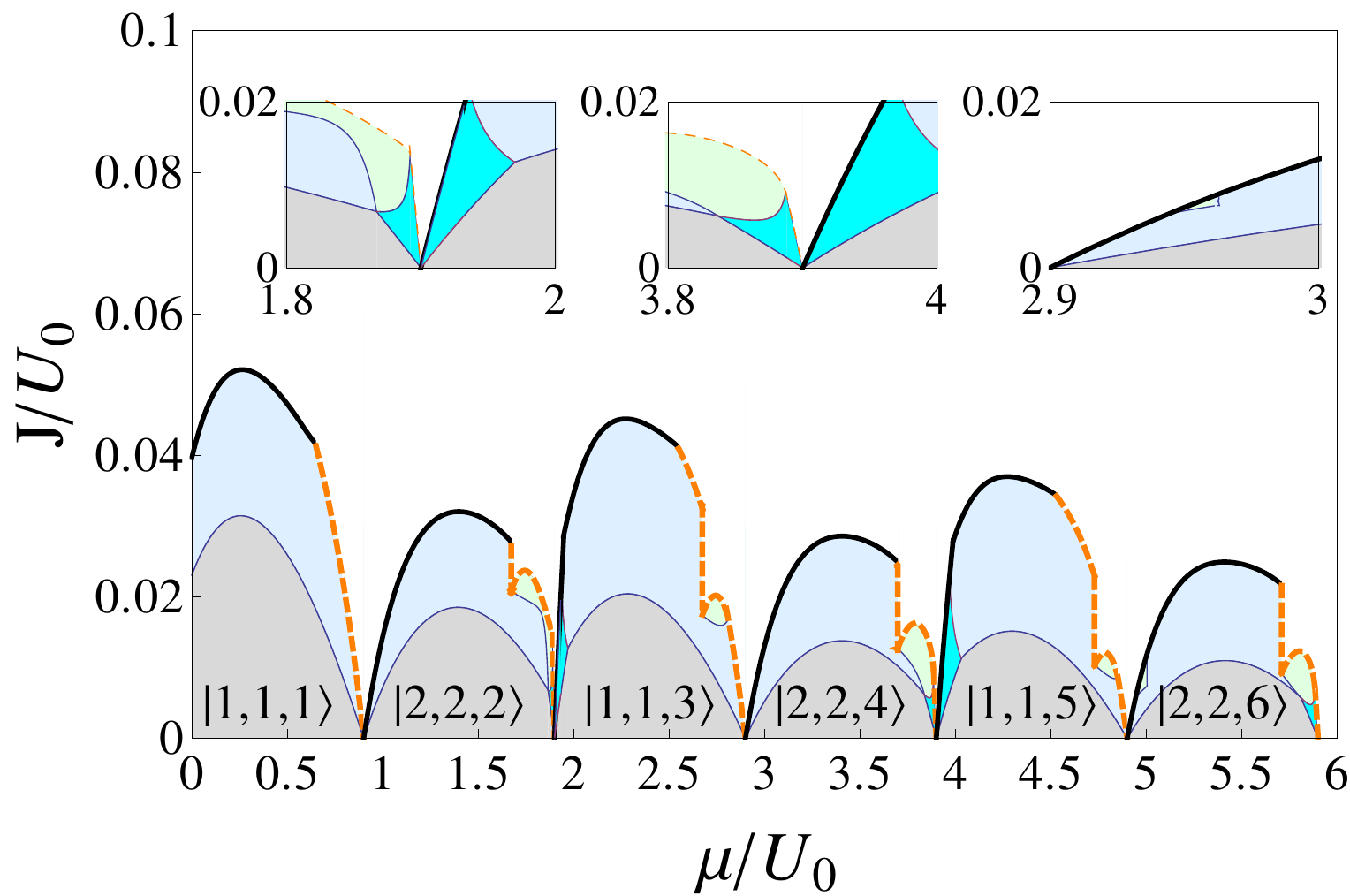}}\hfill{}
\subfloat[\label{fig:51f}$U_{2}/U_{0}=0.15$.]{\raggedright{}
\includegraphics[width=7.5cm]{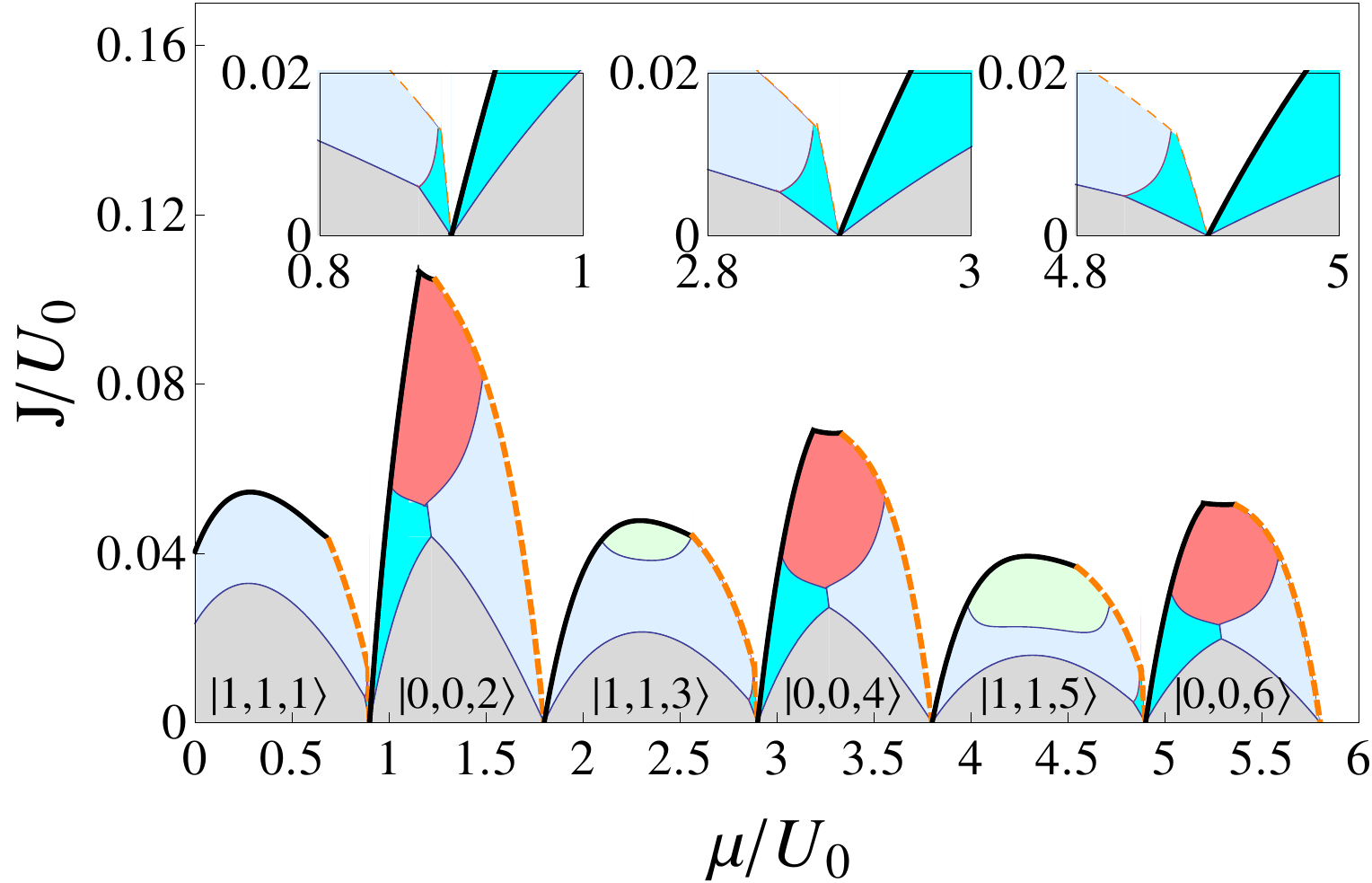}}

\caption{Superfluid phases calculated analytically and numerically for different
values of spin-dependent interaction strength $U_{2}/U_{0}$ and at external field $\eta/U_{0}=0.2$:
$\Psi_{1}\neq0,\:\Psi_{0}=\Psi_{-1}=0$ (blue); $\Psi_{1}\neq0,\:\Psi_{-1}\neq0,\,\Psi_{0}\neq0$ (red);
$\Psi_{-1}\neq0,\:\Psi_{0}=\Psi_{1}=0$ (cyan); and $\Psi_{1}\neq0,\:\Psi_{-1}\neq0,\,\Psi_{0}=0$ (green).
The validity ranges (\ref{eq:5-273-1-1}) and (\ref{eq:5-274-1}) correspond to the black and dashed
orange lines, respectively. Moreover, Mott lobes are characterized by  gray color.}
\label{fig:Superfluid-phases1-1}
\end{figure}

Extremizing the effective potential (\ref{eq:18}) according to (\ref{eq:4-22}) we find at first the
location of the quantum phase transition 
\begin{eqnarray}
zJ_{c}={\rm{min}\atop \alpha}\frac{1}{a_{2}^{(0)}(\alpha,0)},\label{eq:4-25-1}
\end{eqnarray}
which turns out to coincide with the mean-field result in Ref.~\cite{key-6,key-60-1,key-15}, see \fref{fig:Superfluid-phases1-1}.
Moreover, inserting (\ref{eq:18}) into (\ref{eq:4-22}) yields also the different superfluid phases
for ferromagnetic and anti-ferromagnetic interactions with and without magnetization at zero temperature.
If there is more than one solution, we must take the one which minimizes the effective potential for
some system parameter. At first, we observe that the condensate in the superfluid phase above the first
Mott lobe shows, indeed, a sharp increase \cite{key-3}. Thus, the condensate density can not be valid
deep in the superfluid phase. Therefore, it is necessary to determine the range of validity of the Ginzburg-Landau
theory. To this end, we use the fact that we can not have more particles in the condensate than we have
in total. This leads to the condition that the sum over the condensate densities $\sum_{\alpha}\left|\Psi_{\alpha}\right|^{2}$
is equal to the average number of particles per lattice site $\left\langle n\right\rangle =\left.-\frac{\partial\Gamma}{\partial\mu}\right|_{\Psi=\Psi_{\rm{eq}}},$
i.e. 
\begin{eqnarray}
\sum_{\alpha}\left|\Psi_{\alpha}\right|^{2} & = & \left\langle n\right\rangle ,\label{eq:5-273-1-1}
\end{eqnarray}
which is graphically shown in \fref{fig:Superfluid-phases1-1}. But we read off from \fref{fig:Superfluid-phases1-1}
that condition (\ref{eq:5-273-1-1}) breaks down at the end of the lower Mott lobes. There we have to
use an additional criterion to obtain a finite range of validity. To this end we complement condition
(\ref{eq:5-273-1-1}) by the additional ad-hoc restriction that above Mott lobe $n$ the condensate density
can not be larger than $n+1$, yielding the boundary 
\begin{eqnarray}
\sum_{\alpha}\left|\Psi_{\alpha}\right|^{2} & =n+1,\label{eq:5-274-1}
\end{eqnarray}
which is depicted in \fref{fig:Superfluid-phases1-1} as a dashed orange line.

Now we show at zero temperature that our Ginzburg-Landau theory distinguishes between various superfluid phases for a ferromagnetic and anti-ferromagnetic interaction with and
without external magnetic field within the validity range of our theory. Without external magnetization
the superfluid phase is a polar (ferromagnetic) state for anti-ferromagnetic (ferromagnetic) interactions,
which is characterized by $\Psi_{1}\neq0,\,\Psi_{-1}=\Psi_{0}=0$ $\left(\Psi_{0}\neq0,\,\Psi_{-1}=\Psi_{1}=0\right)$,
in accordance with previous mean-field results \cite{key-6,key-29}. In the presence of the magnetic field  the phase diagram
does not change for the ferromagnetic interaction as the minimization of the energy implies the maximum
of spin value as it is in the case without $\eta$ except the degeneracy with respect to $m$ is lifted,
so the ground state becomes $\left|n,n,n\right\rangle .$ For an anti-ferromagnetic interaction, however,
the situation is more complicated with an external magnetic field due to two competing effects. Whereas,
the anti-ferromagnetic interaction energetically favors anti-parallel spins, the external magnetic field
has the tendency to align the spins. In the following we show that this competition leads to the appearance
of different superfluid phases with different magnetic properties. 

In \fref{fig:Superfluid-phases1-1}, we study the predictions of the Ginzburg-Landau theory on how
the external magnetic field affects the superfluid phases in the case of an anti-ferromagnetic interaction,
i.e. $U_{2}>0$. In this context it is important to determine the degeneracy when two states have the
same energy with equal particle number but different total spin. Using the ground state energy (\ref{eq:3})
and the degeneracy condition $E_{S,S,n}^{(0)}=E_{S+2,S+2,n}^{(0)}$, we obtain the critical spin-dependent
interaction strength $U_{2}^{\mathrm{crit}}=\eta/\left(S+\frac{3}{2}\right)$ at the external magnetic
field $\eta$ \cite{key-15,key-16,key-10}. With this, we get the resulting phase diagrams below and
above the critical spin-dependent interaction strength: 

When $U_{2}/U_{0}$ is 0.02, the superfluid phase becomes $\Psi_{1}\neq0,\,\Psi_{-1}=\Psi_{0}=0$ where
the ground state $\left|n,n,n\right\rangle $ is the state with maximum spin for all six lobes as shown
in \fref{fig:51a}. Above the first critical value $U_{2\rm{even}}^{(1)}/U_{0}=0.036$ both
the spin $S$ and the magnetic $m$ quantum numbers change from $\left|6,6,n\right\rangle $ to $\left|4,4,n\right\rangle $
for even lobes as shown in \fref{fig:51b}. We remark that if the spin-dependent interaction increases,
the effect of the external magnetic field decreases, so the Mott lobes increase. The phases $\Psi_{1}\neq0,\,\Psi_{-1}\neq0,\,\Psi_{0}=0$
and $\Psi_{-1}\neq0,\,\Psi_{1}=\Psi_{0}=0$ appear in the SF phase for the fifth and sixth lobes. We
note that the phase $\Psi_{-1}\neq0,\,\Psi_{1}\neq0,\,\Psi_{0}=0$ appears twice in the sixth lobe. The
right phase in the sixth lobe results from the change of $S$ and $m$ for the seventh lobe from $\left|7,7,7\right\rangle $
to $\left|5,5,7\right\rangle $, which happens at the critical value $U_{2\rm{odd}}^{(1)}/U_{0}=0.0308$. 

Beyond the critical value $U_{2\rm{odd}}^{(2)}/U_{0}=0.044$ the values of $S$ and $m$ for the
odd lobes change from $\left|5,5,n\right\rangle $ to $\left|3,3,n\right\rangle $ as shown in \fref{fig:51c}.
Similarly, the phases $\Psi_{1}\neq0,\,\Psi_{-1}\neq0,\,\Psi_{0}=0$ and $\Psi_{-1}\neq0,\,\Psi_{1}=\Psi_{0}=0$
appear in the SF phase for the fourth and fifth lobe and the phase $\Psi_{1}\neq0,\,\Psi_{-1}\neq0,\,\Psi_{0}=0$
increases in the sixth lobe. After the critical value $U_{2\rm{even}}^{(2)}/U_{0}=0.05714$, the
ground states for the even lobes change from $\left|4,4,n\right\rangle $ to $\left|2,2,n\right\rangle $
as shown in \fref{fig:51d}. By the same way the phases $\Psi_{1}\neq0,\,\Psi_{-1}\neq0,\,\Psi_{0}=0$
and $\Psi_{-1}\neq0,\,\Psi_{1}=\Psi_{0}=0$ are seen in the SF phase for the third and fourth lobe. 

When $U_{2}$ increases beyond the critical value $U_{2\rm{odd}}^{(3)}/U_{0}=0.08$, the
ground states for the odd lobes change from $\left|3,3,n\right\rangle $ to $\left|1,1,n\right\rangle $
as shown in  \fref{fig:51e}. The phases $\Psi_{1}\neq0,\,\Psi_{-1}\neq0,\,\Psi_{0}=0$ and $\Psi_{-1}\neq0,\,\Psi_{1}=\Psi_{0}=0$
appear in the SF phase for the even and odd lobes. After the critical value $U_{2\rm{even}}^{(3)}/U_{0}=0.133$,
$S$ and $m$ change from $\left|2,2,n\right\rangle $ to $\left|0,0,n\right\rangle $ for the even lobes
as shown in  \fref{fig:51f}. Furthermore, the effect of magnetic field becomes very weak because
the value of $\eta$ is close to $U_{2}$. Additionally, spin pairs are produced to get the minimal energy
and, thus, the ground state becomes $\left|0,0,n\right\rangle $ for an even $n$, and $\left|1,1,n\right\rangle $
for an odd $n$. We found the new phase $\Psi_{-1}\neq0,\,\Psi_{1}\neq0,\,\Psi_{0}\neq0$ above  the even
lobes.

Furthermore, inspecting the energies of the respective phases in the vicinity of their boundaries allows
to determine the order of the quantum phase transition. With this we find that the quantum phase transition
from the Mott insulator to the superfluid phase is of second order for spin-1 bosons in a cubic optical
lattice under the effect of the magnetic field at zero temperature. Thus, our finding disagrees with
Kimura et al. \cite{key-61-1} where a first-order SF-MI phase transition was found at a part of the
phase boundary by using the Gutzwiller variational approach. Finally, we observe that the transitions
between the different superfluid phases can be of both first and second order above the same Mott lobe.
For instance, the transition from $\Psi_{1}\neq0,\:\Psi_{0}=\Psi_{-1}=0$ to $\Psi_{-1}\neq0,\:\Psi_{1}=\Psi_{0}=0$
or vice versa is of first order, whereas the transition from $\Psi_{1}\neq0,\:\Psi_{0}=\Psi_{-1}=0$
to $\Psi_{1}\neq0,\:\Psi_{-1}\neq0,\,\Psi_{0}=0$ or $\Psi_{1}\neq0,\:\Psi_{-1}\neq0,\,\Psi_{0}\neq0$
phases or vice versa is of second order.

In conclusion, we have worked out a Ginzburg-Landau theory for spin-1 bosons in a cubic optical lattice
within its range of validity and investigated analytically and numerically at zero temperature the different
superfluid phases for an anti-ferromagnetic interaction in the presence of an external magnetic field. Depending on the particle number, the spin-dependent interaction and the value 
of the magnetic field we find superfluid phases with a macroscopic occupation of the two spin states $\pm 1$ or even of all three spin states 0, $\pm 1$. This is  different from the mean-field approximation which only predicted two superfluid phases 
with spins aligned or opposite to the field direction \cite{key-60-1,key-15}.
It would be interesting to study how these results would change in a frustrated triangular optical lattice \cite{key-53}  or in a superlattice \cite{key-6x}.
\ack
We acknowledge discussions with Antun Balaz and Mathias Ohliger as well as  financial support from both the Egyptian Government  and  
the German Research Foundation (DFG) via the Collaborative Research Center SFB/TR49 Condensed
Matter Systems with Variable Many-Body Interactions.

\end{document}